\documentclass[preprint,showpacs,showkeys,preprintnumbers,amsmath,amssymb]{revtex4}

\usepackage{graphicx}% Include figure files
\usepackage{dcolumn}% Align table columns on decimal point
\usepackage{bm}% bold math

\begin{document}

\title{Oscillating Solitons Pinned to a Nonmagnetic Impurity in Layered Antiferromagnets}% Force line breaks with \\

\author{L. A. S. M\'ol}
\author{A. R. Pereira}
 \thanks{Corresponding author : Afranio Rodrigues Pereira.\\
Tel.: 55-31-3899-2988;    fax: 55-31-3899-2483.}
 \email{apereira@mail.ufv.br}
\affiliation{Departamento de F\'{\i}sica,Universidade Federal de Vi\c{c}osa,36571-000,
Vi\c{c}osa, Minas Gerais, Brazil}

\author{W. A. Moura-Melo}
\affiliation{Departamento de Ci\^{e}ncias B\'{a}sicas, Faculdades Federais Integradas de Diamantina, Diamantina, 39100-000, Minas Gerais, Brazil.}
\date{\today}

\begin{abstract}
We argue that an oscillatory motion of impurity-pinned solitons may occur in layered antiferromagnetic compounds. The characteristic frequencies of these modes, that may be detected by resonance or inelastic neutron scattering, are estimated analytically and depend on the soliton sizes and types.
\end{abstract}



\maketitle

In recent years there has been much interest in the static and dynamic properties of layered compounds. These materials  are well-known to exhibit quasi-two-dimensional (2D) magnetism at low temperatures and, in addition to the possibilities in technological applications, they have been useful in testing theories pertinent to low dimensional systems. The 2D behavior of these structures arises because the ratio of the interlayer coupling $J_z$ to intralayer coupling $J$ is of the order $J_z/J \approx 10^{-3} - 10^{-5}$. The 2D effects can be experimentally observed in a narrow temperature interval just above the ordering temperature. The theory of these systems is contained in the Heisenberg Hamiltonians (isotropic and anisotropic). The isotropic Heisenberg magnets in two dimensions are described by the Hamiltonian $H=\pm J \sum_{i,j} \vec{S}_i \cdot \vec{S}_j$, where $\vec{S}_i$ is the spin located at site $i$, the summation is taken over the nearest-neighbor sites on a square lattice, and the - and + signals represent ferromagnetic and antiferromagnetic coupling constants  respectively. The continuum version of this Hamiltonian is given by the 2D $O(3)$ non-linear $\sigma$ model,  which has long been of interest also to particle theorists because of its strong analogy with four-dimensional Yang-Mills theory. For Heisenberg antiferromagnet (AFM), the non-linear $\sigma$ model is described in terms of the N\'{e}el state and gives the antiferromagnetic long-wavelength fluctuations in the limit of zero temperature. It is represented by the following Hamiltonian :
\begin{equation}
H_\sigma=\frac{1}{2} J \int [(\partial_\mu \vec{n}) \cdot (\partial^\mu \vec{n})]d^2x, \quad \mu=1,2,
\end{equation}
where $\vec{n}$ denotes the N\'{e}el unit vector $(n_1^2 +n^2_2 + n_3^2=1)$ characterizing the direction of the sublattice magnetization of the antiferromagnet. It is also well-known that this model supports interesting soliton-like solutions, obtained by Belavin and Polyakov\cite{1} using topological considerations. In terms of the conformal representation $w=(n_1+in_2)/(1-n_3)$, which is simply the stereographic projection of the N\'{e}el vector onto the complex plane, the Belavin-Polyakov soliton ({\it BV}-soliton) located at the origin and with unitary topological charge is expressed as $w_s=z/R$, where $z=x+iy$ and $R$ is the soliton size. It implies that the third N\'{e}el vector component $n_3$ is given by
\begin{equation}
n_3=\frac{|z|^2-R^2}{|z|^2+R^2}.
\end{equation}
Furthermore, it is useful to change variables from $(x,y)$ to $(n_3,\phi=\tan^{-1} (y/x))$, in terms of which we can express the ratio $n_2/n_1=\tan \phi$. Qualitatively, the configuration is radial, pointing down at the origin and up at infinity. The energy of these excitations is found to be independent of the soliton size due to the scale invariance of the non-linear $\sigma$  model and it is given by $E_s=4 \pi J$. The significance of these non-linear excitations for 2D magnets was recognized early in connection with the critical properties of these systems. For example, it was shown that the existence of large localized excitations will cause the correlation length to remain finite at any nonzero temperature\cite{1}. Considering the dynamical properties, the translation motion of this structure may be responsible for a central peak in the out-of-plane dynamic correlation function\cite{2,3}. Recently, Zaspel {\it et al}\cite{4,5} have shown that the Belavin-Polyakov solitons dominate the thermodynamics in the fluctuation region immediately above the N\'{e}el temperature of a large class of nearly classical 2D antiferromagnets. Besides, these authors\cite{6,7} have also shown that the introduction of a very small amount of nonmagnetic impurities into the magnetic sites of a classical 2D antiferromagnet creates a new type of static (impurity-pinned) soliton that affects the Arrhenius, $e^{-E_s/T}$, temperature-dependent EPR linewidth by drastically changing the parameter $E_s$. Using an approach on a discrete lattice, they considered the spin vacancy located at the soliton center and obtained two different impurity-solitons ( the {\it P}-soliton and the {\it I}-soliton), which were detected in the layered antiferromagnet $(C_3H_7NH_3)_2M_xMn_{l-x}Cl_4$ through EPR measurements\cite{6,7}. The {\it P}-soliton has the same structure of the {\it BV}-soliton but without a spin at its center, which results in a vortex singularity at the center. The {\it I}-soliton has smaller size and energy than the {\it P}-soliton, so it also has a configuration slightly different from the {\it P}-soliton. In this letter we argue that these solitons oscillate on the static spin vacancy and estimate the frequencies of these oscillatory modes of solitons analytically.  It is suggested that this effect may be observed in electron spin resonance or inelastic neutron scattering experiments.

In a recent paper, Pereira and Pires\cite{8} considered the problem of a soliton interacting with a nonmagnetic impurity in the continuum limit by modifying Hamiltonian (1) introducing a nonmagnetic impurity potential as follows:
\begin{equation}
H_I=\frac{1}{2} J \int [(\partial_\mu \vec{n}) \cdot (\partial^\mu \vec{n})]V(\vec{r})d^2x, \quad \mu=1,2,
\end{equation}
where $V(\vec{r})=1$ if $|\vec{r}-\vec{d}|\geq a$, and $V(\vec{r})=0$ if $|\vec{r}-\vec{d}|< a$. Here, the impurity is localized at position $\vec{d}$ from the origin, and $a$ is the lattice constant. This lack of magnetic interaction inside the circle of radius $a$, means that a spin located at $\vec{d}$ was removed from the lattice. From now we will consider the nonmagnetic impurity localized at the origin, $\vec{d}=(0,0)$. If the soliton center coincides with the impurity center, this field theory reproduces the two soliton solutions of Refs. \cite{6,7} and still gives the sizes, energies and configurations of these solitons in an analytical way. For the {\it P}-soliton, almost entire space is mapped onto the  plane, but part of the spin space sphere around the origin is not included in the mapping. It means that the stereographic projection of the N\'{e}el vector onto the complex plane is perfect for $r=|z|\geq a$, so we can write the {\it P}-soliton configuration as
\begin{equation}
\frac{n_{2P}}{n_{1P}}=\tan{\phi}; \quad n_{3P}=\left\{\begin{array}{rc}
0 & \mbox{for} \quad r<a, \\
n_3 & \mbox{for} \quad r \geq a.
\end{array}\right.
\end{equation}
For the {\it I}-soliton, the mapping is smoothly deformed, leading to the following configuration
\begin{equation}
\frac{n_{2I}}{n_{1I}}=\tan{\phi}; \quad n_{3I}=\left\{\begin{array}{rc}
0 & \mbox{for} \quad r<a, \\
n_3+b(R)(1-n_3^2) & \mbox{for} \quad r \geq a,
\end{array}\right.
\end{equation}
where the function $b(R)$ was obtained minimizing the energy of excitation (5) for each value of the soliton size and it is given in the interval $0<b(R) \leq 1/2$\cite{8}. The subscript {\it P} and {\it I} refer to the topological and impurity-centered solitons ({\it P}- and {\it I}-solitons) respectively. The energy of these two solitons depends on $R$ with the {\it I}-soliton energy smaller than the {\it P}-soliton energy, but as $R \to \infty$, $E_I \to E_P \to 4 \pi J$. This continuum theory permits also to compare the solitons size $(R_I=[(1+4b^2)^{1/2}-2b]R, \quad R_P=R)$ showing that the {\it I}-soliton is smaller than the {\it P}-soliton\cite{8}.

The field theory given by (3) still shows that, if the soliton center is displaced in relation to the nonmagnetic impurity center, an effective interaction between the soliton and impurity that is attractive at short range is  induced.  If the soliton center and the spin vacancy center are separated by a distance $r_0$, the system energy can be approximated by\cite{8}
\begin{equation}
E_{r_0>a}\cong E_s \left[ 1- \frac{1}{4}\left(\frac{Ra}{r_0^2+R^2}\right)^2+F_-(r_0,R)-F_+(r_0,R) \right],
\end{equation}
where
\begin{equation}
F_\pm (r_0,R)=\frac{R^2a(r_0 \pm a)^2}{[(r_0 \pm a)^2+r_0^2]^{1/2}[(r_0 \pm a)^4 - R^4]}.
\end{equation}
Equation (6) shows that, for $r_0<R$, the soliton energy decreases as the soliton center approximates to the impurity. It indicates that there is a short-range attractive well with length scale on the order of $R$. On the other hand, if the soliton center is localized at distance $r_0>R$ from the spin vacancy, the effective potential between the soliton and the nonmagnetic impurity is repulsive. However, although expression (6) correctly suggests that it is energetically favorable for the soliton to nucleate about the spin vacancy, this equation is valid only for separations larger than one lattice spacing $(r_0>a)$\cite{8}. In fact, in the limit that the separation vanishes $(r_0 \to 0)$, the soliton configuration changes to either the {\it P}-soliton or {\it I}-soliton configuration, which are radially symmetric and represent local minima of energy\cite{8}. Then, if there is a spin vacancy exactly at the soliton center,  Eq.(6) should give the values $E_I(R)$ or $E_P(R)$, depending on which soliton was formed (the {\it I}- or {\it P}-soliton). Hence, in the limit $r_0 \to 0$, we use an expansion of the energy around the bottom of the well as follows
\begin{equation}
E_{r_0 \to 0}(r_0,R)\cong E_\alpha (R) + \frac{1}{2\!}r_0^2 \left(\frac{d^2E}{dr_0^2}\right)_{r_0=0}+ \frac{1}{3\!}r_0^3 \left(\frac{d^3E}{dr_0^3}\right)_{r_0=0}+ \ldots ,
\end{equation}
where $\alpha=${\it P} or {\it I} represents one of the two possible solitons. For small displacements, we can neglect the high-order terms in the expansion. The complete function $E(r_0,R)$ is likely to be a smooth interpolation between the limit (8) and Eq. (6), with the transition between these expressions occurring at $r_0 \approx a$. Keeping only the harmonic term in (8), we get
\begin{equation}
\left(\frac{d^2E}{dr_0^2}\right)_{r_0=0}=K=2 \frac{E_s-E_\alpha (R)}{a^2} + 2 E_s \left[ \frac{R^2}{\sqrt{5} [R^4-(2a)^4]}- \frac{R^2}{4 (R^2+a^2)^2}\right],
\end{equation}
where $R>2a$. Equation (8) is interpreted as follows: the first term is just the rest mass of the {\it P}- or {\it I}-soliton and the remaining terms may be attributed to the interaction potential energy $U(r_0)$ of the impurity-soliton system, as a function of their separation $r_0$. This attractive interaction at short-range suggests that, in general, soliton solutions near nonmagnetic impurities may not be static. Spinless atoms may exert some force on the magnetic solitons, in addition to distorting them, and cause them to accelerate. If the force obtained from a potential agrees with the mass times acceleration of the soliton, then, due to the form of the effective potential in Eq.(8), which is an attractive well, we expect that the soliton center may have an oscillatory motion on the impurity. This Newtonian behavior may really happen to magnetic nonlinear excitations. For example, in 2D easy-plane magnetic systems, Wysin\cite{9,10} has shown that vortex excitations have  motions that are  the same as those predicted from a simple Newtonian dynamical equation of motion for the vortex center, involving a force $\vec{F}$ and effective mass $M$, i.e.,
\begin{equation}
\vec{F}=M \frac{d \vec{\nu}}{dt},
\end{equation}
where $\vec{\nu}$ is the velocity of the vortex center. In this case, the mass depends on the system size since a vortex is not a localized object. But, if the anisotropy is weak, the mass of the out-of-plane AFM vortices  may actually reach a finite limit for large system size. Besides, it was shown that for a simple harmonic motion, the mass was found as
\begin{equation}
M=\frac{K}{\omega_0^2}
\end{equation}
where $\omega_0$ is the translation mode frequency and $K$ is the spring constant\cite{9,10}.  Here,  we assume  that the results of Ref.\cite{9} extrapolate, with appropriate modifications, to the isotropic case, i.e., to the Lorentz-invariant {\it O}(3) nonlinear $\sigma$ model, which represents the isotropic AFM. Thus, since the soliton is a localized structure, the rest mass would be expressed as $M_\alpha=E_\alpha(R)/c^2$\cite{11}, where $c=2JSa/ \hbar$ is the velocity of the long-wavelength spin waves ($\alpha=${\it P} or {\it I}).

As it can be seen from Eq. (8), a small displacement $r_0$ of the {\it P}- or {\it I}-soliton center away from the exact impurity center (which is, in our model, the center of the circle of radius $a$) involves an energy increase. It is clear that these off-centered solutions are true bound states, because their energy is smaller than the free soliton mass. Any perturbation of the soliton will make it drift to the center of the well. The off-centered solutions are unstable. In fact, for displacements much less than the lattice spacing, the short-range potential well is close to harmonic, and can be described by an effective linear force acting on the soliton. The effective force constant $K=(d^2E/dr_0^2)_{r_0=0}$ can be approximated by Eq. (9). Then, the frequencies of the oscillations of the solitons in a trapped state may be estimated analytically. Since this elastic constant is a function of $E_\alpha(R)$, then, to determine the frequencies of the solitons oscillation modes, an analytical expression for the rest energy of the pinned-solitons would be useful. In Ref.\cite{6}, it was shown that the energies of the two solitons can be written through the same expression $E_\alpha(R)=E_sR^2/(R^2+a_\alpha^2)$, but with different $a_\alpha$'s ($a_p=0.23a$ for the {\it P} and $a_I=1.01a$ for the {\it I} soliton). Using these results, the solitons oscillation mode frequencies on a nonmagnetic impurity can be estimated as
\begin{equation}
\omega_\alpha=\sqrt{2}\frac{c}{a}\left[\frac{a_\alpha^2}{R^2}+(R^2+a_\alpha^2)\left(\frac{a^2}{\sqrt{5} [R^4-(2a)^4]}- \frac{a^2}{4 (R^2+a^2)^2}\right)\right]^{1/2}.
\end{equation}
Figure 1 shows the frequencies for the two solitons as a function of their sizes. Of course, large solitons oscillate slower than small solitons. As a consequence, in general, {\it P}-solitons oscillate slower than {\it I}-solitons.

The analysis of the spectrum of small-disturbance (the mesons of the theory or magnons) on a non-linear excitation always exhibits zero-frequency modes $\omega=0$, whenever the theory possesses a continuum symmetry. Although the soliton's energies (6) (or (8)) are invariant under spatial rotations around the z-axis (which for the hedgehog is equivalent to an iso-rotation around the internal 3-axis), they are not invariant under a translation in the x-y plane. In the presence of a nonmagnetic impurity, there is no possibility of a soliton to move away from a position by translations and keeping $E$ unchanged. The impurity  breaks the translational symmetry of the system, and thus, there are not zero-frequency modes associated to translations on the xy-plane (translation zero-modes). The oscillatory modes given by Eq.(12) can be considered to be driven by the force caused by the lattice defect (the spin vacancy), which can be evaluated independently of the spin-wave spectrum. To see this, we consider the change of the out-of-plane component of the N\'{e}el vector $n_3$ as the soliton center moves for a small displacement $r_0<<a$ from the impurity center. Since the soliton oscillates on the nonmagnetic impurity, we write $r_0=A \sin (\omega_\alpha t)$, where A is the oscillation amplitude assumed to be small. Then, the time dependent difference $\epsilon_\alpha (\vec{r},t)=n_{3\alpha}(r_0)-n_{3\alpha}(0)$ gives the change of the out-of-plane component of the N\'{e}el vector as the soliton ({\it P} or {\it I}) oscillates. This difference can be easily evaluated. After linearizing in the small quantity, we obtain $\epsilon_\alpha (\vec{r},t) \approx Im [A^\mu \partial_\mu n_{3\alpha} \exp(-i \omega_\alpha t)]$. This expression shows that, if $\omega_\alpha$ were zero, the function $n_{3\alpha}(A^\mu)=n_{3\alpha}(0)+A^\mu \partial_\mu n_{3\alpha}$ would describe a soliton which had been uniformly translated on the plane by a small amount $A^\mu=(A^1,A^2)$ (translational mode). However, our solution has a nonzero frequency $\omega_\alpha$, which implies in $n_{3\alpha}(A^\mu)=n_{3\alpha}(0)+A^\mu \partial_\mu n_{3\alpha}\exp(-i \omega_\alpha t)$ leading to an oscillatory motion for the soliton. These results can have a different interpretation. Here, the translational modes can be identified by viewing the motions of the spins that result when a given spin-wave eigenfunction is added to the original soliton structure. Then, the small-disturbance waveform $\epsilon_\alpha (\vec{r},t) = A^\mu \partial_\mu n_{3\alpha} \exp(-i \omega_\alpha t)$ may also be viewed  as out-of-plane spin waves with well defined frequency and angular momentum about the static soliton structure $n_{3\alpha}$, which the center is localized exactly at the impurity center. It means that, instead of thinking about an oscillating soliton, we can imagine a stopped soliton surrounded by these magnons. In fact, using polar coordinates $(r,\phi)$ and supposing that $A^\mu=(0,A)$, we have
\begin{equation}
\epsilon_\alpha (\vec{r},t)=Im\left(\frac{\partial n_{3\alpha}}{\partial y} e^{-i \omega_\alpha t}\right)=Im\left(\frac{4AR^2r}{(r^2+R^2)^2} e^{ i(m\phi-\omega_\alpha t)}\right)
\end{equation}
where $m= \pm 1$ is the angular momentum channel. In this interpretation, the above expression gives the time dependent deformation of the static soliton. In particular, if one could observe the spin motions around the impurity (or in the central core of the soliton), their instantaneous directions could be used to estimate the position of the soliton center. Equation (13) shows  that the small-disturbance amplitude goes to zero as $r \to \infty$ and hence, it describes magnon modes, which are localized at a soliton. The soliton-impurity interaction may induce  magnon local modes around the spin vacancy. These magnon local modes on the nonmagnetic impurity have frequencies that depend on the soliton type ({\it P} or {\it I} ) and size. In a discrete lattice, the possible sizes $R$ are given by $R_l=la$ ($l$ is a positive integer) and then, the possible modes have  also a discrete  spectrum, whose frequencies are
\begin{equation}
\omega_{\alpha,l}=\sqrt{2}\frac{c}{a}\left[\frac{\delta_\alpha^2}{l^2}+(l^2+\delta_\alpha^2)\left(\frac{1}{\sqrt{5} (l^4-16)}- \frac{1}{4 (l^2+1)^2}\right)\right]^{1/2}.
\end{equation}
where $\delta_\alpha$ is a number ($\delta_p=0.23$ for the {\it P}-soliton and $\delta_I=1.01$ for the {\it I}-soliton). Note that $l \geq 3$, since the vacancy at the soliton center forbids the formation of solitons with size smaller than $3a$ in a discrete lattice. For $l=3$, $\omega_{P,3}=0.302c/a$ and $\omega_{I,3}=0.561c/a$, while for $l=4$, $\omega_{P,4}=0.197c/a$ and $\omega_{I,4}=0.402c/a$, which typically have values $(10^{11} - 10^{13})s^{-1}$. It is expected that local modes with well defined frequencies are induced around nonmagnetic impurities via soliton. Resonances at the characteristic frequencies can be in principle observed in electron spin resonance or inelastic neutron scattering experiments. Since these modes can occur only if {\it P}- or {\it I}- solitons are centered on the nonmagnetic impurity, resonance and neutron scattering measurements provide some methods to experimentally detect this effect and consequently, these two pinned solitons.

Equation (12) keeps certain similarities with the results obtained by Wysin\cite{9,10} for vortices in easy-plane magnets. The spin configuration of a vortex (or the magnetization inhomogeneity) extends to the limits of the system, causing its inertia against motion inside the system to depend on the system size. Hence, the effective vortex mass increases as the square of the system radius and the translational mode frequencies diminish as the reciprocal system size\cite{9,10}. In our case, since the magnetization inhomogeneity depends on the soliton radius, the translational frequencies diminish as the reciprocal soliton size.

The above results are valid only for small displacements satisfying $r_0<<a$.  For large displacements, in which the soliton has to move for more than one lattice spacing, $r_0>a$, nonlinear terms are present in the interaction force and  besides, in  real magnetic materials, discreteness effects must be considered. These effects can be described through a periodic pinning potential. These calculations are out of the scope of this paper.

Layered magnetic compounds of the form $(C_nH_{2n+1}NH_3)_2MX_4$, where $M$ is a transition metal ion and $X$ is $Cl$ or $Br$, are nearly-classical spin systems for $M=Mn^{2+}$ and $X=Cl$ (the $Mn$ ion has a spin 5/2). Hence, under nonmagnectic doping ions (such as $Mg$ or $Cd$), these materials are good candidates to experimentally detect these modes and testing the theory.  Finally, we would like to point out that our theory may also have some relevance to quantum spin systems\cite{12,13,14}. In fact, the well known technique of spin coherent states effectively replaces spin operators by classical vectors $\vec{S}=(\sin \theta \cos \phi,\sin \theta \sin \phi, \cos \theta)$ and incorporates the quantum features by means of the path integral over all space-time configurations of $(\theta, \phi)$. Important 2D systems exhibiting quantum magnetism are the high-T$_c$ cuprate superconductors. In particular, concerning the problem of impurities, recent nuclear magnetic resonance measurements involving the high-T$_c$ cuprate superconductors, have shown that a substitution of the spin-1/2 $Cu^{2+}$ in the $Cu-O$ plane by a strong nonmagnetic impurity, such as spin-0 $Zn^{2+}$,  enhances the antiferromagnetic correlations around the impurity\cite{13}. We also suggest that these results have some relevance in the quantum Hall effect, where spin textures called skyrmions play important roles in the two-dimensional electron gas\cite{14,15}.

\begin{acknowledgments}
This work was partially supported by CNPq (Conselho Nacional para o Desenvolvimento da Pesquisa) Brazil.
\end{acknowledgments}

\newpage

\begin{figure*}
\includegraphics[height=5cm, keepaspectratio]{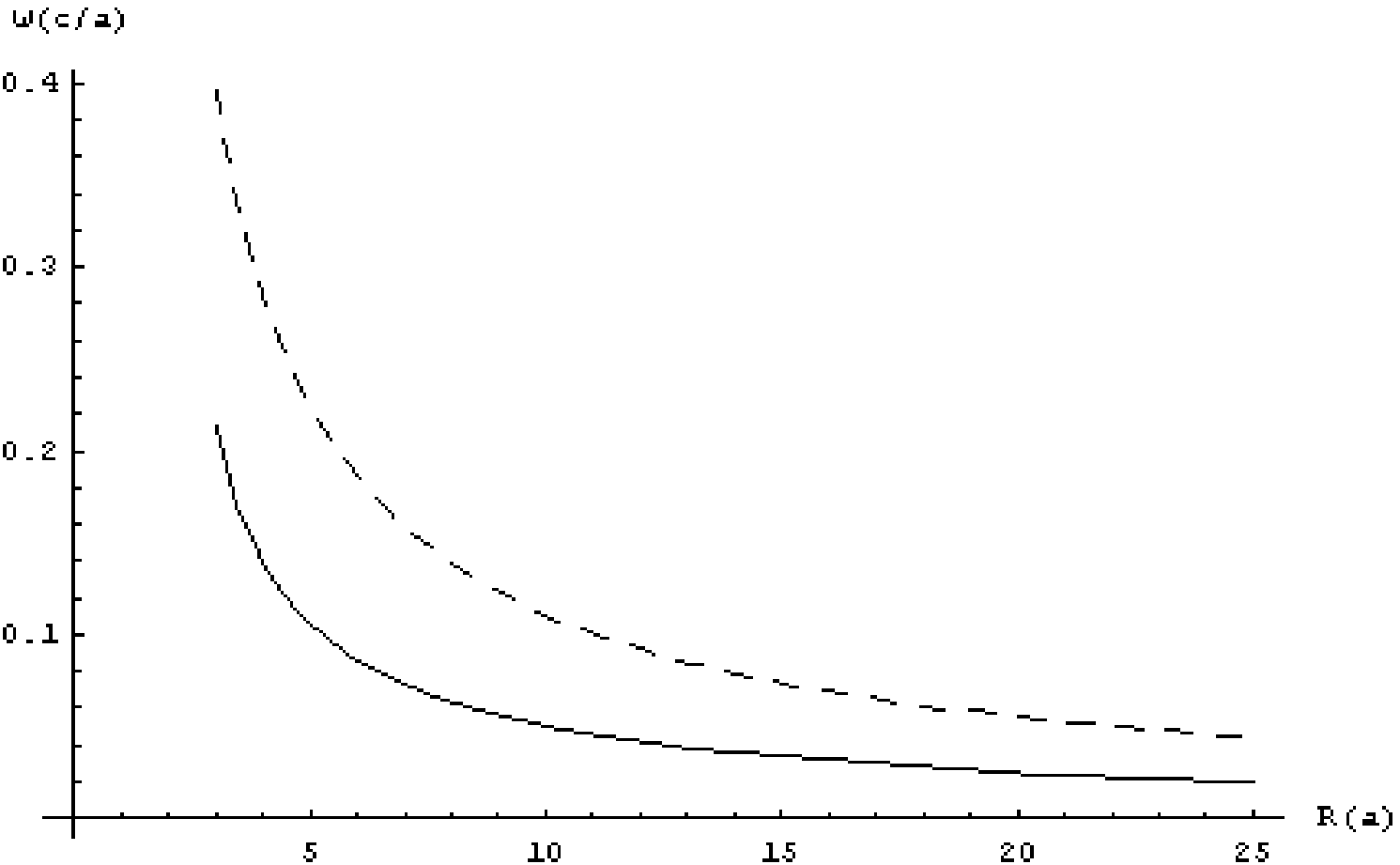}
\caption{Oscillation frequencies as a function of the soliton size. Solid curve is for P-solitons and dashed curve is for I-solitons.}
\end{figure*}

\end{document}